\documentclass[aps,prl,twocolumn,superscriptaddress]{revtex4}

\usepackage{graphicx}
\usepackage{dcolumn}
\usepackage{epstopdf}
\usepackage{amssymb,amsmath}

\usepackage{bm}
\usepackage{color}

\newcommand{\PZT}{Pb(Zr$_{1-x}$Ti$_x$)O$_3$~}

\newcommand{\PMNPT}{$_{(1-x)}$Pb(Mg$_{1/3}$Nb$_{2/3}$)O$_3$-$_x$PbTiO$_3$~}

\begin{document}                  
\widetext
\leftline{Version 1.6 as of \today}

\title{Improving the Functional Control of Aged Ferroelectrics using Insights from Atomistic Modelling}

\author{J. B. J. Chapman}
\affiliation{Department of Physics and Astronomy, University College London, Gower Street, London WC1E 6BT, UK}
\affiliation{National Physical Laboratory, Hampton Road, Teddington, TW11 0LW, UK}
\author{R. E. Cohen}
\affiliation{Department of Physics and Astronomy, University College London, Gower Street, London WC1E 6BT, UK}
\affiliation{Geophysical Laboratory, Carnegie Institution of Washington, Washington, DC 20015, USA}
\affiliation{Ludwig Maximilian University of Munich, 80539 M\"{u}nchen, Germany}
\author{A. V. Kimmel}
\affiliation{Department of Physics and Astronomy, University College London, Gower Street, London WC1E 6BT, UK}
\author{D. M. Duffy}
\affiliation{Department of Physics and Astronomy, University College London, Gower Street, London WC1E 6BT, UK}
\date{\today}


\begin{abstract}
We provide a fundamental insight into the microscopic mechanisms of the ageing processes.  Using large scale molecular dynamics simulations of the prototypical ferroelectric material PbTiO$_3$, we demonstrate that the experimentally observed ageing phenomena can be reproduced from intrinsic interactions of defect-dipoles related to dopant-vacancy associates, even in the absence of extrinsic effects. We show that variation of the dopant concentration modifies the material's hysteretic response. We identify a universal method to reduce loss and tune the electromechanical properties of inexpensive ceramics for efficient technologies.

\end{abstract}

\maketitle      
                

Technologies utilising ferroelectric components are ubiquitous in modern devices, being used from mobile phones, diesel engine drive injectors and sonar to print heads and non-volatile memory \cite{Genenko2015,Glaum2012,Scott1989,Catalan2012}. Doping with transition-metals has been shown experimentally to improve electromechanical properties of widely used ferroelectrics. For example, doping of BaTiO$_3$, PbTiO$_3$, \PZT (PZT) and \PMNPT (PMNPT) is used to improve the functional properties and efficiency of these simple and cheap oxides \cite{Neumann1987,Arlt1988,Ren1997,Ren2000,Ren2004,Zhang2005,Zhang2008,Wu2008,Perez2011}. However, the fundamental origin of the electromechanical improvements is not understood and requires full characterisation to enable properties to be directly tuned for purpose and functional lifetimes to be accurately predicted. 

Dopant interactions can be classified as intrinsic (bulk/volume) or extrinsic (boundary). Extrinsic coupling is associated with domain wall and grain boundary effects. Defects, including dopants and vacancies, migrate to domain walls and subsequently pin their propagation, resulting in fatigue of the material's switching properties \cite{Carl1978}. Intrinsic effects occur independent of interaction with domain walls, as they arise due to the interaction between defect induced dipoles $\vec{p}_d$ and the spontaneous polarisation of the domain surrounding the defect site $\vec{P}_s$. Strong evidence from electron paramagnetic spin resonance (ESR) and density functional theory (DFT) calculations has shown dopants/impurities, such as iron Fe$^{3+}$ and copper Cu$^{2+}$, in PZT (or Mn$^{2+}$ in BaTiO$_3$) substitute the B-cations as acceptors, which bind to charge compensating oxygen vacancies $V_O^{2-}$ to form thermodynamically stable defect complexes \cite{Erhart2007,Eichel2008}. In Kr\"{o}ger-Vink notation, the divalent dopant-vacancy associates can be written as ($B^{''}_{Ti}+V^{\bullet \bullet}_O)^{\times}$, where $B^{''}_{Ti}$ is an unspecified divalent dopant subsituting a Ti$^{4+}$ site, $V_O^{\bullet \bullet}$ is an oxygen vacancy with a +2e charge relative to the defect free site, $'$ identifies a negative charge unit (-e), $\bullet$ represents a positive charge unit (+e) and $\times$ stands for charge neutrality. Density functional theory calculations have shown defect-dipoles $\vec{p}_d$ spontaneously form for ($Fe^{'}_{Ti}+V^{\bullet \bullet}_O)^{\bullet}$ \cite{Mestric2005} and ($Cu^{''}_{Ti}+V^{\bullet \bullet}_O)^{\times}$ \cite{Eichel2008} associates in PbTiO$_3$ and ($Mn^{''}_{Ti}+V^{\bullet \bullet}_O)^{\times}$ in BaTiO$_3$ \cite{Zhang2005,Nossa2015}, and the energetically favourable oriention is along the polar axis [001]. Group-IIIB and group-VB acceptor substitutes on Ti sites in PbTiO$_3$ have been shown to form immobile clusters of dopant-vacancy associates which have different structures when the associate is aligned parallel or perpendicular to the polar axis \cite{Zhang2008b}. ($V_{Pb}^{''}+V_O^{\bullet \bullet})^{\times}$ divacancy complexes in PbTiO$_3$ have been calculated to have a local dipole moment twice the bulk value \cite{Cockayne2004}.  

Ageing is simply defined as the change in a material's properties over time. It has been proposed that in aged ferroelectrics, defect-dipoles produced from dopant-vacancy associates will slowly rotate to align in parallel with the domain symmetry to minimise its energy state \cite{Lambeck1986,Ren2004,Kimmel2012}. The co-alignment and subsequent correlated behaviour of these aged defect-dipoles has been proposed to create a macroscopically measurable internal bias, which in turn has been conjectured to be responsible for experimentally observed ageing phenomena, including a 10-40 fold increase in piezoelectric coefficients, shifts in the hysteresis along the electric field axis and pinched/double hysteresis loops typically associated with antiferroelectrics \cite{Genenko2015,Ren2004,Zhang2005,Zhang2008}.

\begin{figure*}
\centering
    \includegraphics[width=0.9\textwidth]{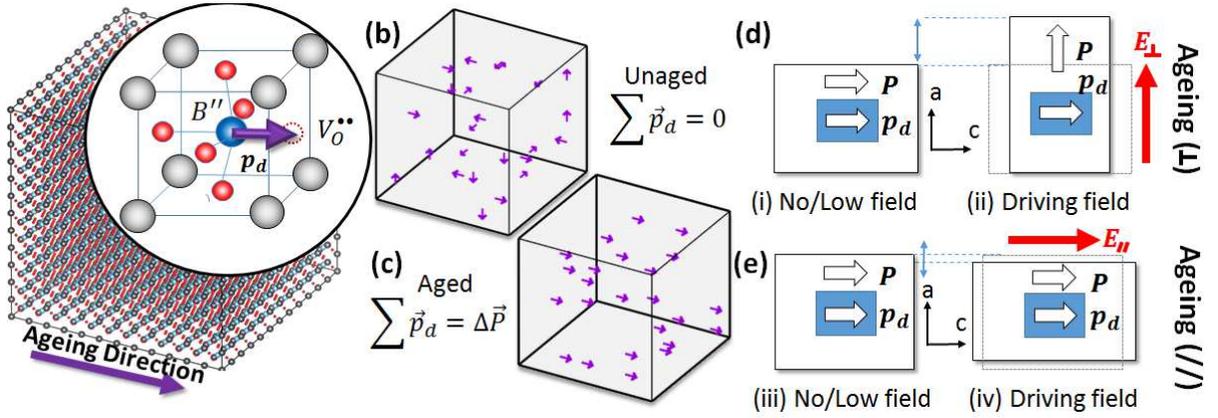}
    \caption{System configuration for MD simulations of ageing. (a) PbTiO$_3$ supercell including defect-dipoles from ($B^{''}_{Ti}+V^{\bullet \bullet}_O)^{\times}$ associates (inset). (b) Defect-dipoles in unaged PbTiO$_3$ are randomly orientated. (c) Aged PbTiO$_3$ is modelled by aligning all defect dipoles along the ageing direction [100]. (d) Schematic of perpendicular ageing. A [001] poling field is applied perpendicular to the ageing orientation. (e) Schematic of parallel ageing. The poling field is applied parallel to the ageing orientation.
    }
    \label{fig:setup}
\end{figure*}

In this letter, we use large scale classical molecular dynamics to model ageing arising from defect-dipoles of dopant-vacancy associates in tetragonal bulk lead titanate (PbTiO$_3$). We show that all the experimentally observed large signal effects (P-E and S-E hysteresis) of aged prototype perovskite ferroelectrics; pinched and double hysteresis, shifted hysteresis and a large recoverable electromechanical response can be reproduced from intrinsic effects alone and we identify the microscopic mechanisms of each case.


We study ideal and aliovalent-doped bulk PbTiO$_3$ using classical molecular dynamics (MD) as  implemented in the DL$\_$POLY code \cite{Dlpoly}. We use the adiabatic core-shell interatomic potentials derived in Gindele \emph{et al} \cite{Gindele2015} that reproduces the properties of bulk and thin films of PbTiO$_3$ in excellent agreement with DFT calculations \cite{Gindele2015,Chapman2017}. The prototype PbTiO$_3$ has been chosen as it has a single ferroelectric phase, which reduces competing effects and because it is a parent compound for two of the most widely used ferroelectric materials in industry (PZT/PMNPT).

In this study we investigate volume effects, therefore, three-dimension periodic boundary conditions are implemented to mimic an infinite crystal, devoid of surfaces, interfaces and grain boundaries. We choose a moderate supercell constructed from $12\times 12 \times 12$ unit cells, approximately 125~nm$^3$, corresponding to 8,640 atoms (for the ideal bulk). This system size is large enough for ensemble sampling but sufficiently small to prevent the formation of 90$^{\circ}$ domain walls. We use the Smooth Particle Mesh Ewald (SPME) summation for the calculation of Coulomb interactions. Coupling between strain and polarisation is enabled using the constant-stress Nos\'{e}-Hoover ($N\sigma T$) ensemble with thermostat and barostat relaxation times of 0.01~ps and 0.1~ps, respectively. A 0.2~fs timestep is used in all instances. Initial calculations were run at 100~K to prevent diffusion of the vacancies \cite{Morozov2010,Smyth1994} and to allow the correct characterisation of each effect. The temperature dependence, for the range from 50~K to 400~K, is then investigated.

We calculate polarisation - electric field (P-E) hysteresis using a quasistatic approach. Starting at 0~kV/mm, the electric field is cycled between the limits $\pm150$~kV/mm in 16.7~kV/mm intervals. For each field strength the system is restarted using the coordinates, velocities and forces from the previous calculation and equilibrated for 4~ps to enable the system to equilibrate following the E-field impulse. This is followed by an 8~ps production run over which statistics are collected (total of 12~ps per iteration). We calculate the local polarisation by considering conventional Ti-centred unit cells as implemented in references \cite{Sepliarsky2011,Gindele2015,Chapman2017}. Further details are provided in the Supplementary Information. 

A dopant-vacancy concentration $n_d=100(N_{Ti}^{ideal}-N_{B^{''}})/N_{Ti}^{ideal}$ is introduced into the supercell initially containing $N_{Ti}^{ideal}$ Ti atoms, by randomly selecting a total of  $N_{B^{''}}$ Ti atoms to be replaced with generic divalent dopants $B_{Ti}^{''}$. Each dopant is coordinated by six nearest neighbouring oxygen-sites from which a charge compensating oxygen vacancy, $V_O^{\bullet \bullet}$, can be introduced. This configuration mimics ($B^{''}_{Ti}+V^{\bullet \bullet}_O)^{\times}$ dopant-vacancy associates observed from ESR experiments (Figure~\ref{fig:setup}a).
In experiments it is observed that the properties of an aged sample can be removed by heating above the Curie temperature for a long period and then rapidly quenching. It has been hypothesised that during this `un-ageing' process in the cubic phase of the prototype ferroelectric, each orientation of the defect-dipole is equally probable such that vacancies will thermally hop between the neighbouring oxygen site adjacent to the dopant and eventually 1/6 defect-dipoles will populate each of the six possible directions \cite{Ren2004}. These are then frozen when quenched into the ferroelectric phase. Ferroelectrics can then be intentionally aged again by applying a bias field for a significantly long period. It is believed this causes defect-dipoles to align. Even in the absence of an ageing field, defect-dipoles in a sample left for a long period will align with the spontaneous polarisation of the domain \cite{Zhang2008}. When constructing the supercell for a particular simulation, the choice of which oxygen is removed neighbouring the dopant depends on the aged/unaged condition:

(1) Unaged condition.  To simulate unaged tetragonal PbTiO$_3$ we assign $N_{B^{''}}/6$ defect-dipoles along each of the six possible orientations causing the total moment to cancel, Figure~\ref{fig:setup}b.

(2) Aged condition.  To simulate an aged PbTiO$_3$ sample, each $V_O^{\bullet \bullet}$ is selected to situate on the oxygen-site along the ageing direction (defined below) relative to its associated dopant. For these simulations we arbitrarily choose the ageing direction along $+\hat{x}$ (see Figure~\ref{fig:setup}a). This initialises all defect dipoles $\vec{p}_d$ as parallel, polarised along [$\bar{1}00$] as shown in Figure~\ref{fig:setup}c.

The ageing direction is defined relative to the driving field for the hysteresis characterisation. If the defect dipoles are co-aligned with the driving field we label this as aged($\parallel$) (Figure~\ref{fig:setup}d), whereas perpendicular alignments are labelled aged($\perp$) (see Figure~\ref{fig:setup}e). The strain is calculated as $\Delta \epsilon=(c_0-c)/c$ where $c_0$ is the relaxed lattice constant (parallel to the drive field orientation) under no applied field. Further details of the computational methodology are provided in the Supplementary Information.

\begin{figure}[ht]
\centering
	\includegraphics[width=0.45\textwidth]{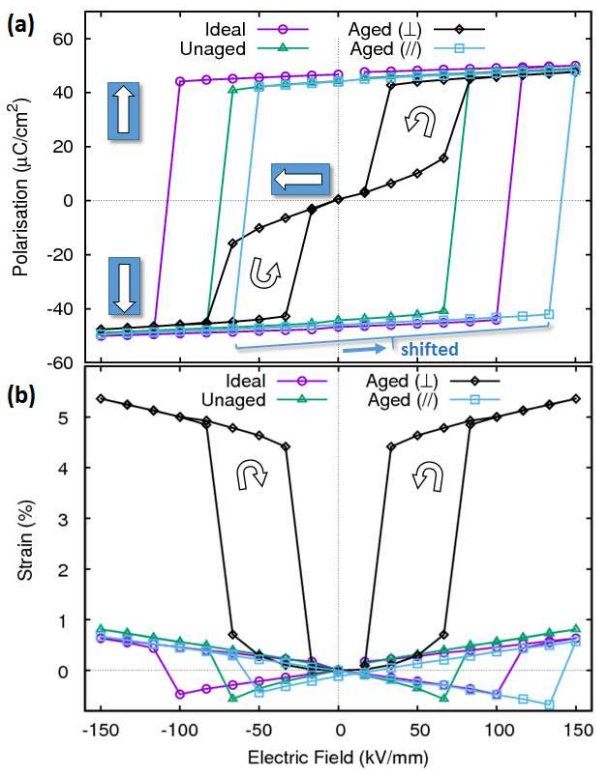}
	\caption{Hysteresis of doped PbTiO$_3$ when defect free (purple-circles), unaged (green-triangles), poled parallel to the aged orientation (blue-squares); and aged perpendicular to the poling direction (black-diamonds). (a) Polarisation - Electric field hysteresis. (b) Electrostrain hysteresis.}
	\label{fig:ageing}
\end{figure}


Firstly we discuss results obtained at 100~K, to observe the ideal behaviour without thermal diffusion or hopping. The calculated P-E hysteresis of PbTiO$_3$ in response to an external driving field is shown in Figure~\ref{fig:ageing}a for the defect-free bulk, unaged and the two aged conditions. In all instances, the response is highly non-linear, typical of ferroelectrics. For the ideal bulk case a symmetric, square loop indicative of a hard ferroelectric is observed. We note our bulk coercive field $E_c^{int}$ corresponds to the material's intrinsic coercive field, which greatly exceeds those measured experimentally for Pb-based ferroelectrics \cite{Glaum2012}. This is because our model excludes grain boundaries, surfaces and domain walls which would all act as nucleation sites, which lower the energy barrier for reversal in physical samples. Our result of 130~kV/mm matches other MD models \cite{Zeng2011} and is in excellent agreement with the intrinsic coercive field of 150~kV/mm calculated using density functional perturbation theory \cite{Sai2002}.

Figure~\ref{fig:ageing}a shows the hysteresis of an aged single domain simulated sample, with a defect concentration of 1.38\%, in response to a driving field perpendicular to the direction in which the material was aged. Interestingly, when the system is equilibrated with no applied field the spontaneous polarisation $\vec{P}$ reorientates parallel to the ageing direction (See Supplementary Figure 2 and cartoon schematic in Figure~\ref{fig:setup}d). This shows the internal bias created from the defect-dipoles is sufficient to overcome the switching barrier \cite{Kimmel2012}. This observation provides direct evidence supporting the work of Zhang \emph{et al} \cite{Zhang2008} who observed that non-switching defect-dipoles from (Mg$^{''}_{Ti}-$V$_O^{\bullet \bullet})^{\times}$ associates in BaTiO$_3$ create restoring forces that promote reversible domain switching. Under the application of the perpendicular driving field there is an almost linear response until 67~kV/mm ($\approx E_c^{int}/2$), at which point the field strength is sufficient to switch the polarisation parallel to the drive field. As the electric field decreases to zero, the polarisation again reorientates along the ageing axis such that no remnant polarisation $P_r$ remains in the poling direction. Thus in our work, the iconic double-hysteresis indicative of aged ferroelectrics is observed without the requirement of either domain walls or grain boundaries \cite{Ren2004,Genenko2015}.

When poling parallel to the ageing orientation (Figure~\ref{fig:setup}e) the system exhibits a shifted hysteresis curve along the electric-field axis as shown in Figure~\ref{fig:ageing}a. Such an effect is well documented in the literature when there is a preferred orientation of the defect dipoles in the poling direction~\cite{Rojac2016,Carl1978,Lambeck1986}. 

In the unaged simulation we observe a symmetric square E-P hysteresis loop (Figure~\ref{fig:ageing}a). The computed coercive field of the unaged PbTiO$_3$ is reduced relative to the ideal bulk value by 35\% ($0.65E_C^{int}$). The reduction of the coercive field from $E_c^{int}$ occurs because the dopant-vacancy associates break local symmetry creating localised areas where the activation energy for nucleation of reverse domains is reduced \cite{Vopsaroiu2010}. 

The electrostrain (S-E) of each condition is shown in Figure~\ref{fig:ageing}b. Symmetric butterfly S-E curves are observed for both the ideal ($n_d=0$\%) and unaged ($n_d=$1.38\%) simulations, in excellent agreement with unaged ferroelectrics measured by experiment \cite{Glaum2012,Balke2009,Zhang2004}. In order to test the validity of results in comparison to experiment, we calculate the d$_{33}$ piezoelectric tensor coefficient of bulk PbTiO$_3$ using a fluctuation-perturbation theory approach \cite{Garcia1998} and second derivative matrices using the GULP package \cite{Gulp} (see Supplementary Information for further details). We find $d_{33}=47\pm1$~pC/N from the fluctuations, and 49~pC/N from the derivatives, which agree with each other and agree well with values measured on polycrystalline PbTiO$_3$ films (52-65~pC/N) \cite{Kighelman2002}.

Ageing parallel to the poling field is shown to induce an asymmetric S-E hysteresis (Figure~\ref{fig:ageing}b). Instances of large asymmetric S-E loops have been experimentally reported in a range of ferroelectric materials \cite{Tan2014}. In \cite{ShiFan2015}, the authors report a strain difference of 0.15\% in Li doped (Bi$_{0.5}$Na$_{0.4}$K$_{0.1}$)$_{0.98}$Ce$_{0.2}$TiO$_3$ ceramics which they propose is due to alignment of (Li$^{'''}_{Ti}$-V$_O^{\bullet \bullet})^{'}$ associates. Using our prototypical system, we provide evidence that the asymmetry is likely to arise from an excess orientation along the poling direction and is a general feature of ageing that may be exploited for technological applications.

It has been observed that ageing of doped BaTiO$_3$ is capable of producing a large recoverable non-linear electric field induced strain of 0.75\%; far greater than those measured in PZT or PMNPT \cite{Ren2004,Zhang2005}. It is argued that a strong restoring force from aligned defect-dipoles enables polar axis rotation parallel to the defect-dipoles enabling reversible switching of 90$^{\circ}$ domains and could lead to the realisation of strain values of 6\% in PbTiO$_3$. In Figure~\ref{fig:ageing}b, we show that ageing perpendicular to a poling field in PbTiO$_3$ leads to a large recoverable strain in excess of 4.5\% (black-diamonds). This large non-linear strain arises from the reorientation of the polar axis along the ageing direction due to the internal bias from the defect dipoles at subswitching fields (see Supplementary Figure 2. $c\to a$, $\Delta \epsilon = (c_0-a)/a$). Observing the switching behaviour, we find that, in this instance, the 90$^{\circ}$ switching occurs via near-homogeneous polarisation rotation over a small field range rather than nucleation and growth of 90$^{\circ}$ domains - a switching mechanism predicted in bulk PbTiO$_3$ \cite{Zeng2011} and BaTiO$_3$ \cite{Kimmel2012}. Therefore, we further show the volume effect of the dopant-vacancy associates to be the fundamental cause of this ageing phenomenon and a domain wall mechanism is not required for the full reproduction of experimental observations. Our ageing results are in excellent agreement with a complementary bond valence model study of ageing in ideal BaTiO$_3$ using fixed dipoles introduced into the crystal structure \cite{Liu2017}.

\begin{figure}[ht]
\centering
	\includegraphics[width=0.45\textwidth]{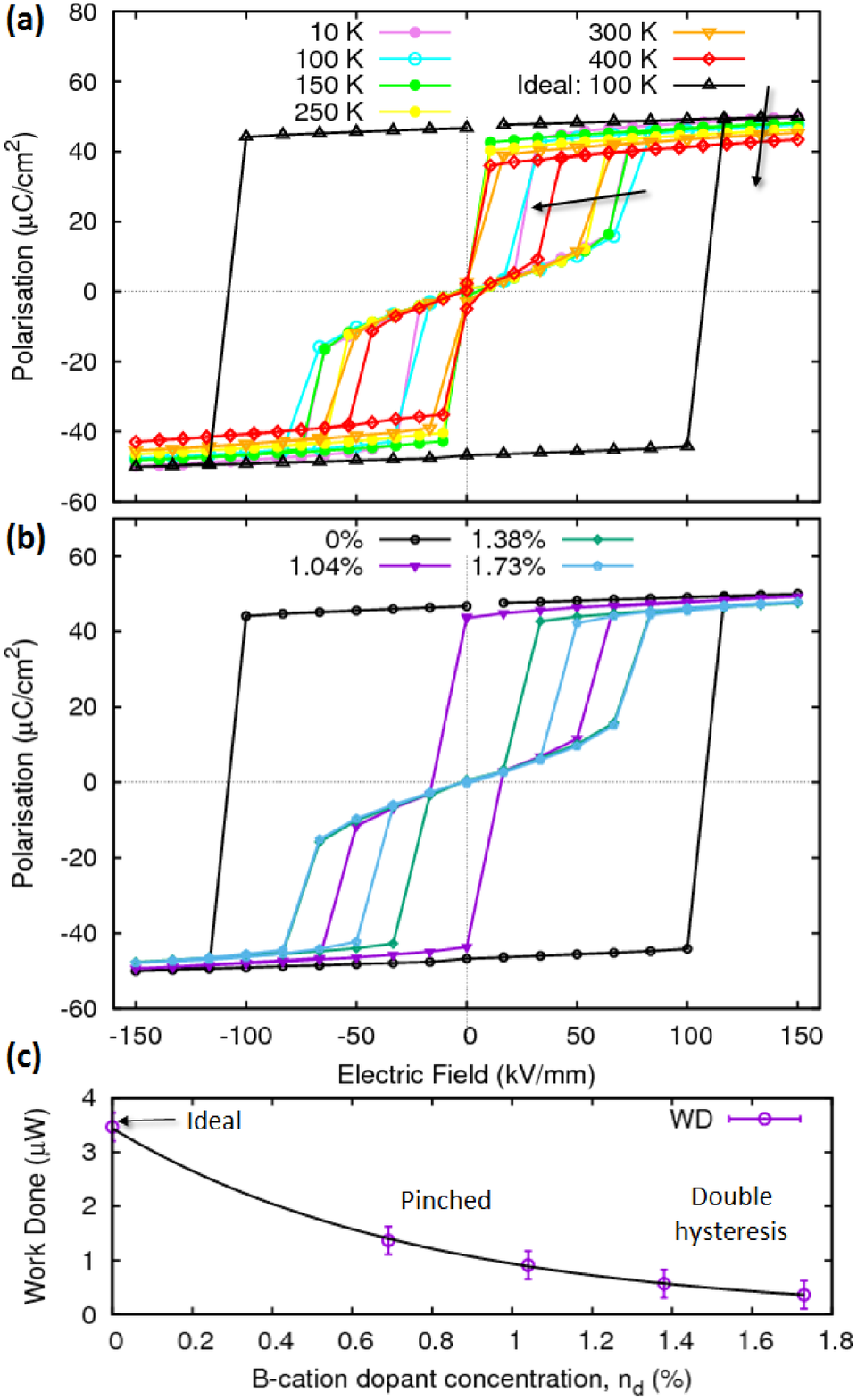}
	\caption{Aged hysteresis properties. (a) Temperature dependence of the P-E hysteresis for $n_d$=1.38\%. (b) The effect of dopant concentration on the hysteresis of PbTiO$_3$. Low concentrations retain the square loop of the pure ferroelectric. Pinching is observed at dopant concentrations greater than 0.78\% which close to form double hysteresis loops with further increases in concentration ($\approx$~1.38\%). (c) Work done to create hysteresis over a period $T$ (area enclosed in the loop $W=\frac{1}{T}\oint EdP$). A solid line is plotted to guide the eye.}
	\label{fig:concentration}
\end{figure}

The results of an investigation into the effect of temperature on the ageing phenomenon in PbTiO$_3$ is shown in Figure~\ref{fig:concentration}a for $n_d=1.38$\%. As the temperature increases, a decrease in the effective coercive field and saturation polarisation are observed, corresponding to a narrowing of the double hysteresis as indicated by the trend arrows. This is analogous to the behaviour known for the square loop of the ideal prototype. Near room temperature under poling fields comparable to $E_c$, vacancy hopping becomes thermally activated causing limited events whereby a subset of defect dipoles reorientate. This was observed by tracking the displacement of each oxygen atom relative to its initial position. This reorientation can create asymmetric loops as seen at 400~K ($0.67T_c$), clearly demonstrating that at high temperatures/large fields the defect-dipoles can readily realign, elucidating the microscopic mechanism for aged $\to$ unaged transitions. No defect-dipoles were observed to switch below 300~K (0.5T$_c$). At 300~K a single hopping event was observed (1/24 vacancies) and two (1/12 vacancies) at 400~K, over the full hysteresis. We note that due to the relatively short simulation times these hopping frequencies will be under-sampled for accurate statistics and will be an interesting subject for future investigation.

Defect concentrations close to and above 1.38\% are shown to form closed double hysteresis loops described previously (Fig.~\ref{fig:concentration}b). As the concentration is increased the enclosed area of the hysteresis loops decrease due to the increased strength of the internal bias, which lowers the barrier for the reorientation of the polar axis. For intermediate defect concentrations (0.78\% in this model), we find pinched hysteresis loops are produced. This form of P-E loop is the most common large signal observation noted in experimental studies of aged ferroelectrics \cite{Rojac2016,Glaum2012,Morozov2008}. We find that the work dissipated (area enclosed by the P-E loop) decreases with the dopant level (Fig.~\ref{fig:concentration}c).  Increased defect concentrations start to pinch the square loop which, upon further increases, leads to a closed double hysteresis and gradual reduction of area. Thus, the dissipated energy losses, effective coercive fields and hysteretic behaviour of ferroelectric materials can be controlled by varying the applied fields and dopant levels. We note that in our study we are limited by the constraint of zero total dipole moment in our unaged simulation cell, which restricts the number of dopants $N_{B^{''}}$ to factors of six. Thus, the concentrations identifying pinching and double hysteresis are, in fact, upper bounds.


In conclusion, we use molecular dynamics to model ageing in boundary-free single domain doped PbTiO$_3$. We show that all the large-signal characteristics of ageing: pinched/double hysteresis, hysteresis shifts and large recoverable non-linear strains, can be reproduced from intrinsic effects of defect-dipoles from dopant-vacancy associates alone, resulting from the net defect dipole orientation with respect to the poling field. Varying the concentration of dopants was found to modify the material's hysteretic response, suggesting a mechanism for tuning ferroelectric and electromechanical properties for enhanced device performance. This work identifies and clarifies the microscopic mechanisms involved the ageing phenomena and suggests practical methods to inexpensively improve functional performance of ferroelectric ceramic based technologies. 


Funding was provided by the EPSRC (EP/G036675/1) via the Centre for Doctoral Training in Molecular Modelling and Materials Science at University College London and the National Measurement Office of the UK Department of Business Innovation and Skills. Computer services on Archer were provided via membership of the UK's HPC Materials Chemistry Consortium funded by EPSRC (EP/L000202). We acknowledge the use of the UCL facilities LEGION and GRACE, and computational resources at the London Centre for Nanotechnology. REC acknowledges support of the US Office of Naval Research, the ERC Advanced grant ToMCaT, and the Carnegie Institution for Science.

\bibliography{PTO_Ageing_Bibliography}
\end{document}